\documentclass[aps,prb,amsmath,amssymb,reprint,superscriptaddress]{revtex4-2}

\usepackage{mathrsfs,amsfonts} 
\usepackage{bbm}
\usepackage[colorlinks,citecolor=blue,linkcolor=red,urlcolor=blue]{hyperref}
\usepackage{color}
\usepackage{graphicx}
\usepackage{subfigure}
\usepackage{verbatim}
\usepackage{fourier}
\usepackage{soul}

\begin{document}

\title{Normal and lateral Casimir-Lifshitz forces between a nanoparticle and a graphene grating}

\author{Minggang Luo}
\email[]{minggang.luo@umontpellier.fr}
\affiliation{Laboratoire Charles Coulomb (L2C), UMR 5221 CNRS-Universit\'e de Montpellier, F-34095 Montpellier, France}

\author{Youssef Jeyar}
\affiliation{Laboratoire Charles Coulomb (L2C), UMR 5221 CNRS-Universit\'e de Montpellier, F-34095 Montpellier, France}

\author{Brahim Guizal}
\affiliation{Laboratoire Charles Coulomb (L2C), UMR 5221 CNRS-Universit\'e de Montpellier, F-34095 Montpellier, France}

\author{Mauro Antezza}
\email[]{mauro.antezza@umontpellier.fr}
\affiliation{Laboratoire Charles Coulomb (L2C), UMR 5221 CNRS-Universit\'e de Montpellier, F-34095 Montpellier, France}
\affiliation{Institut Universitaire de France, 1 rue Descartes, Paris Cedex 05, F-75231, France}

\date{\today}
\begin{abstract}
We study the normal and lateral components of the Casimir-Lifshitz (CL) force between a nanoparticle and a 1D graphene grating deposited on a fused silica slab. For this purpose, the scattering matrix approach together with the Fourier modal method augmented with local basis functions are used. 
We find that, by covering a fused silica slab by a graphene grating, the spectrum of the normal CL force at small frequencies is increased by about 100\% for a grating filling fraction of 0.5, and even more when the slab is completely covered. The typically employed additive approximation (the weighted average of the force with and without the graphene coating) cannot provide any information on the lateral CL force, and, as we show, cannot provide an accurate estimation for the normal CL force. When the nanoparticle is laterally shifted ($x_A$), the normal CL force is modulated and remains attractive. On the contrary, the lateral CL force changes sign twice in each period, showing a series of alternating stable and unstable lateral equilibrium positions, occurring in the graphene strips and the grating slits regions, respectively. Finally, we show that the lateral shift effect is sensitive to the geometric factor $d/D$ ($d$ is the separation distance, and $D$ is the grating period). We identify two clear regions: a region ($d/D<1.0$) where the lateral shift significantly affects the CL energy, and a region ($d/D \geq 1.0$) where this effect is negligible, with a crossover at $d\approx D$. Our predictions can have relevant implications to experiments and applications of the CL normal and lateral forces acting on nanoparticles interacting with structured objects at the nano/micro scale, and are also directly valid for atoms close to these nanostructures.
\end{abstract}

\maketitle 

\section{introduction}
{The Casimir-Lifshitz (CL) interaction occurs between polarizable bodies, and originates in quantum and thermal fluctuations of the electromagnetic field \cite{Casimir1948dh,Lifshitz1961}. This interaction, which is a {generalized van der Waals force}, is of great interest for theoretical,  experimental and applicative reasons \cite{Woods2016RevModPhys}. The remarkable dependence of this interaction on the geometry of the interacting bodies has been successfully investigated for several configurations (nanoparticles, spheres, cylinders, gratings...), together with the effect of the material dielectric functions (dielectrics, metals, topological materials...). An extension of the Lifshitz theory, capable of including in a unified way both thermal equilibrium and non-equilibrium configurations for arbitrary objects, is now available \cite{AntezzaPhysRevLett.95.113202,Messina2011PRA,Messina2014}, allowing for the study of both momentum and energy exchange.}

{Among the possible geometries, grating structures are particularly interesting since they are widely used in nanophotonics and can excite high-order diffraction modes, and thus have a significant impact on fluctuation-induced phenomena like CL forces \cite{Hobun,Casimir_Noto,Energy_2_pra,Francesco2013nc,2021prl_Torque} and the radiative heat transfer \cite{Yang2017prl,Messina2017PRB,Hongliang_JHMT}. Materials with different optical properties have {also} been explored. In particular, 2D materials like graphene can show special electromagnetic conductivity features that enable a wide modulation of these effects \cite{Chahine_Multilayer,Grating_JEYAR,LiuPhysRevB.2021,Svetovoy2012prb,Ognjen2012prb,Zheng2017,Volokitin2017Dey,LiuES2022,Lu2022small,Klimchitskaya2022prb,Mauro_Graphene_CLP}. Recently, a two parallel graphene gratings configuration has been investigated to jointly exploit non-trivial geometry and dielectric features for both CL force and heat transfer \cite{Luo2023Casimir_gg,Luo2024shifted}. The radiative heat transfer between a nanoparticle and a graphene grating has also been recently investigated showing remarkable features related to a topological transition from circular to hyperbolic plasmonic modes \cite{Luo2024prb_NFRHT_np_grating}.}

\begin{figure} [htbp]
\centerline {\;\;\;\;\includegraphics[width=0.45\textwidth]{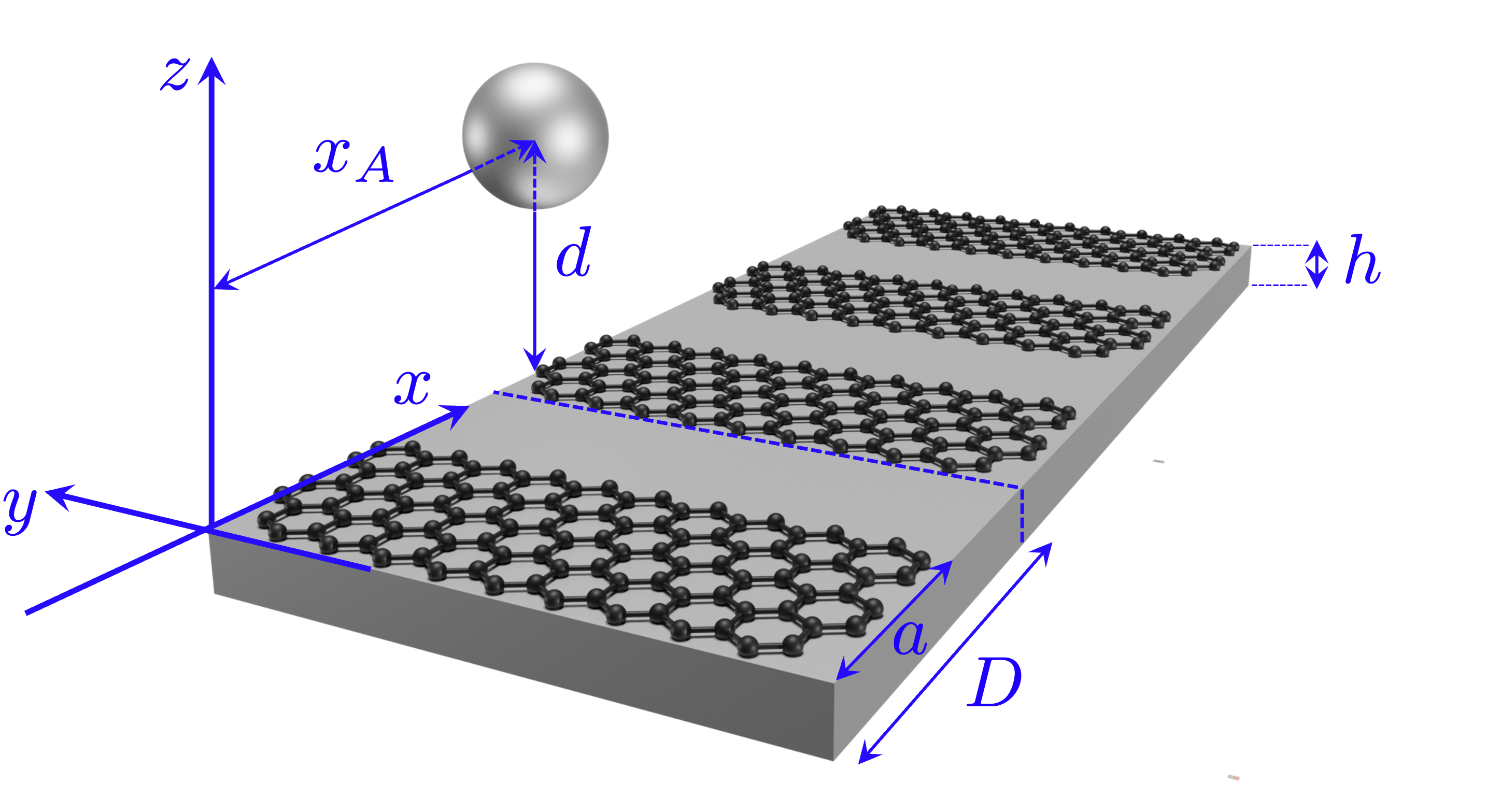}}
\caption{Scheme of the physical system consisting of a nanoparticle whose center is positioned at a distance $d$ from a dielectric slab covered by a strips graphene grating.}
\label{fig:two_gratings_schematic}
\end{figure}

To explore the concurrence of non-trivial geometry and material effects in CL interaction, it is natural to consider a small object that acts as a local probe in close proximity to a graphene-based nanostructured surface \cite{Messina2009PhysRevA,Energy_2_pra,Antezza_2006}. This is what we study in this paper by investigating the CL interaction (energy, normal and lateral forces) for a nanoparticle placed close to a graphene grating, as shown in Fig.~\ref{fig:two_gratings_schematic}.

The paper is structured as follows: In Sec.~\ref{models}, we describe the physical system and the theoretical model. In Sec.~\ref{results_discussion}, we present our numerical results, analyzing the influence of various factors (e.g., filling fraction, chemical potential, and grating period) on the normal and lateral shift-mediated CL interactions. We also study different regimes for the effects related to the lateral shift.

\section{Physical system and theoretical model} \label{models}

{We study the CL interaction between a fused silica (SiO$_2$) nanoparticle of radius $R=20$ nm and a graphene grating coating a planar fused silica slab of thickness $h=20$ nm. The graphene grating has a period $D$, strips of width $a$, a filling fraction $f = a/D$, and a chemical potential $\mu$.  The whole system is at thermal equilibrium, with both bodies and the environment being set at the same temperature $T=300$ K. The nanoparticle center is positioned at $\textbf{R}_A = (\textbf{r}_A, z_A) = (x_A, 0, z_A)$ (we set $y = 0$ due to translation invariance along the strips in the $y$-axis). The separation between the nanoparticle center and the grating is $d = z_A$, and the nanoparticle can be displaced both normally and laterally with respect to the grating plane, as depicted in Fig.~\ref{fig:two_gratings_schematic}.}
 
Assuming that the separation distance between the two bodies is significantly larger than the nanoparticle radius, the CL free energy $E$ of the system, the lateral force $F_x$, and the normal force $F_z$ acting on the nanoparticle are respectively given by \cite{Energy_2_pra,Messina2009PhysRevA}
\begin{eqnarray}
{E}_{}(x_A,d,\mu)   &=&-  \frac{ k_{\rm B} T}{4 \pi^2} \sum_{m=0}^{+\infty}\!\!^{'}   \int_{-\frac{\pi}{D}}^{\frac{\pi}{D}} {\rm d}k_{x}    \int_{-\infty}^{+\infty} {\rm d} k_y  { \rm Tr \left({   \mathcal{R}^{\rm{G}}\mathcal{R}^{\rm{NP}}  } \right)} ,\;\;\;\;\;\;\;\; 
\label{Energy_dipolar}\\
F_x(x_A,d,\mu)  &=&-\frac{\partial {E}_{} (x_A,d,\mu)}{\partial x},
\label{F_x_Der}\\
F_z(x_A,d,\mu)&=& -\frac{\partial {E}_{} (x_A,d,\mu)}{\partial z}.
\label{F_z_Der}
\end{eqnarray}
Here, the sum in Eq.~(\ref{Energy_dipolar}) is carried out over imaginary Matsubara frequencies given by $\xi_m=2\pi m k_{\rm B} T/ \hbar$, where $\hbar$ is the reduced Planck constant,  and $k_{\rm B}$ is the Boltzmann constant. The prime on the sum indicates that the $m=0$ term has to be divided by 2, $k_{x}$ is in the first Brillouin zone $\left[-\frac{\pi}{D},\frac{\pi}{D}\right]$, $k_y \in \mathbb{R}$, and $\mathcal{R}^{\rm{G}}$ and $\mathcal{R}^{\rm{NP}}$ are the reflection operators of the grating and the nanoparticle, respectively, in the (TE,TM) basis. Here, TE and TM represent the two light polarizations: transverse electric and transverse magnetic, respectively.

The operator $\mathcal{R}^{\rm{G}}$ \footnote{corresponds to $\mathcal{R}^{+}$ given by Eq. (55) in \cite{Luo2023Casimir_gg}.} is calculated using the Fourier modal method augmented with local basis functions (FMM-LBF) \cite{PhysRevE.JEYAR,Luo2023Casimir_gg}. This method has been shown to drastically reduce the computation time, however, for the $m=0$ Matsubara frequency, we use the standard Fourier modal method, which ensures more stable integration \cite{Casimir_Noto,Grating_JEYAR}.  At this stage, it is important to note that for each $k_x$ in Eq. \eqref{Energy_dipolar}, and due to the periodicity along the $x$-axis, the operator $\mathcal{R}^{\rm{G}}$ involves all possible channels with parallel wave vectors $k_{xn}=k_x+2\pi n/D$ and normal wave vectors $k_{zn}=\sqrt{k_0^2- k_{n}^2}$, where  $k_n=\sqrt{k_{xn}^2+k_y^2}$, $k_0=\omega/c$, and $n \in \mathbb{Z}$. In the numerical implementation, the integer $n$ will be restricted to the interval $[-N,N]$, where $N$ is the truncation order. It is worth noticing that the truncation order $N$ used in this paper has been set to ensure a relative error of $\lesssim 1\%$.

Concerning the nanoparticle reflection operator $\mathcal{R}^{\rm{NP}}$, it is calculated using the dipole approximation \cite{Mulet2001apl, Joulain2014jqsrt_dipole, ChengQ2023ol}. The dielectric SiO$_2$ nanoparticle is described as an electric dipole moment $\textbf{p}(\omega) = \varepsilon_0 \alpha(\omega) \textbf{E}(\textbf{R}_A, \omega)$, where $\varepsilon_0$ is the vacuum permittivity. The polarizability $\alpha(\omega)$ of the small particle is defined in the Clausius-Mossotti form as $\alpha(\omega) = 4\pi R^3 [\varepsilon(\omega) - 1] / [\varepsilon(\omega) + 2]$, with $\varepsilon(\omega)$ being the relative permittivity of the particle, and $\textbf{E}(\textbf{R}_A, \omega)$ the incident electric field at position $\textbf{R}_A$ at angular frequency $\omega$. Thus, $\mathcal{R}^{\rm{NP}}$ is given by \cite{Messina2011PRA, Messina2014}

\begin{eqnarray}
\left \langle p,{\textbf{k}},n|\mathcal{R}^{\rm{NP}}(\omega)|p',{\textbf{k}}',n' \right\rangle  &= &\frac{i\omega^2\alpha(\omega)}{2c^2k_{zn}} \notag \\ & \times &\left[\hat{\textbf{e}}_p^-(\textbf{k},\omega,n) \cdot \hat{\textbf{e}}_{p'}^+(\textbf{k}',\omega,n') \right] \notag \\ & \times & e^{i \left({k_{xn'}'}-{k_{xn}} \right) {x}_A}  e^{i \left( k_{zn}^{}+k_{z{n'}}' \right)z_A}	,
\label{R2_}
\end{eqnarray}
where the lateral shift ($x_A$) and vertical translation ($d=z_A$) between the {nanoparticle} and the grating have been included through two explicit phase factors.
It is important to note that, here, the definition of the operators for the two interacting bodies is given for real frequencies. However, to calculate the energy and forces in Eqs. \eqref{Energy_dipolar}, \eqref{F_x_Der}, and \eqref{F_z_Der}, it is necessary to obtain $\mathcal{R}^{\rm{NP}}$ and $\mathcal{R}^{\rm{G}}$ at imaginary frequencies by replacing $\omega$ with $i\xi_m$. All previous expressions remain valid if, instead of a nanoparticle, one considers an atom. In that case $\alpha(\omega)$ is to be considered as the atomic polarizability. With the sign convention used here, $F_z<0$ corresponds to an attractive normal force, and $F_x>0$ indicates a lateral force on the nanoparticle directed along the $x$-axis of Fig. \ref{fig:two_gratings_schematic}.

It is worth stressing that the normal force in Eq. \eqref{F_z_Der} can be explicitly expressed as 
\begin{eqnarray}
F_z(x_A,d,\mu)&=&\notag\\
&&\!\!\!\!\!\!\!\!\!\!\!\!\!\!\!\!\!\!\!\!\!\!\!\!\!\!\!\!\!\!\!\!\!\!\!\!\!\!\!\!\!\!\!\!\!\!\!\!\!\!\!\!-\frac{ k_{\rm B} T}{4 \pi^2} \sum_{m=0}^{+\infty}\!\!^{'}  \int_{-\frac{\pi}{D}}^{\frac{\pi}{D}} {\rm d}k_{x}  \int_{-\infty}^{+\infty} \!\!\!{\rm d} k_y   {\rm Tr}\left[ \gamma (\mathcal{R}^{\rm{NP}} \mathcal{R}^{\rm{G}}+\mathcal{R}^{\rm{G}} \mathcal{R}^{\rm{NP}}) \right],\;\;\;\;\;\;\;\;\;
\label{CLF}
\end{eqnarray}
where $\gamma={\rm diag}[{\rm diag}(\kappa_{zn}),{\rm diag}(\kappa_{zn})]$, and $\kappa_{zn}=\sqrt{\xi_m^2/c^2+{{k}}_{n}^2}$. 

In addition, when the filling fraction $f=0$ (bare slab) or $f=1$ (slab fully covered with a graphene sheet), the CL energy and the normal CL force acting on the {nanoparticle} are given by \cite{Lifshitz1961,Messina2014}:
\begin{eqnarray}
{E}_{}(d,\mu)&=&-\frac{ k_{\rm B} T}{4 \pi c^2} \sum_{m=0}^{+\infty}\!\!^{'} \alpha(i\xi_m)   \int_{0}^{+\infty} \frac{Q}{\kappa_z} \mathcal{N} e^{-2\kappa_z z_A} {\rm d} Q, \;\;\;\;\;\;\;\;\;\;
\label{Energy_slab_particle}\\
F_{z} (d,\mu) &=&-\frac{ k_{\rm B} T}{2 \pi c^2 } \sum_{m=0}^{+\infty}\!\!^{'}   \alpha(i\xi_m) \int_{0}^{+\infty} Q  \mathcal{N}   e^{-2\kappa_z z_A}    {\rm d} {Q},\;\;\;\;\;\;\;\;\;\;
\label{CL_Matsubara_slab_particle}
\end{eqnarray}
where  $\mathcal{N} =  \rho_p  ( \xi_m^2+2 c^2 Q^2)-\rho_s  \xi_m^2$, $\kappa_{z}=\sqrt{\xi_m^2/c^2+Q^2}$, $Q$ is the modulus of the in-plane wavevector, $\rho_s$ and $\rho_p$ are the TE and TM Fresnel reflection coefficients for the planar structure [which is a dielectric slab fully coated by a graphene sheet ($f$ = 1) or without graphene ($f$ = 0)] and are explicitly given in Ref. \cite{prb_jeyar2023}.

The electromagnetic properties of graphene are taken into account through its conductivity $\sigma_g$, which depends on the temperature $T$ and the chemical potential $\mu$. This conductivity is the sum of an intraband and an interband contributions \(\left( \sigma_g = \sigma_{\textnormal{intra}} + \sigma_{\textnormal{inter}} \right)\) which, for imaginary frequencies, are given by \cite{Falkovsky2007,Falkovsky2008,Chahine_Multilayer}:
\begin{equation}
\left\{
\begin{array}{rcl}
\begin{aligned}
&\sigma_{\textnormal{intra}} = \frac{2 e^2 k_{\rm B} T}{\pi\hbar^2 (\xi_m + 1/\tau)}\ln\left[ 2 \cosh\left(\dfrac{\mu}{2k_{\rm B}T}\right) \right]      ,
\\&\sigma_{\textnormal{inter}}=    \dfrac{ e^2 \xi_m}{\pi}\int_0^{+\infty} \dfrac{G(x)}{(\hbar \xi_m)^2 + 4x^2} {\rm d}x 	,
\end{aligned}
\end{array}
\right. 
\label{sigma_g}
\end{equation}
where  $e$  is the electron charge, $\tau$ is the relaxation time (we use $\tau = 10^{-13}${s}), $G(x) = \sinh (x/k_B T) / [\cosh (\mu/k_{\rm B} T) + \cosh(x/k_{\rm B} T)]$. As for fused silica, the dielectric response at imaginary frequencies is calculated using the Kramers-Kronig relation, $\varepsilon(i\xi_m)=1+2\pi^{-1}\int_{0}^{\infty} \omega\varepsilon''(\omega)/(\omega^2+\xi_m^2)d\omega$, based on the optical data at real frequencies taken from Ref.~\cite{Palik}.

\section{Results and discussion}
\label{results_discussion}
In this section we show and discuss the frequency spectrum of the CL force, as far as the CL energy, lateral and normal forces.

\subsection{Spectrum of the Casimir-Lifshitz normal force and violation of the additivity approximation}
The Matsubara sum, appearing in the previous expressions of the energy and forces, is made over imaginary frequencies {$i\xi_m$}. Before performing these sums, it is instructive to investigate the behavior of the frequency spectrum of the interaction, i.e. of the function that must be evaluated at each Matsubara frequency.  We consider here the frequency spectrum of the CL normal force. Let us start considering the two simple limiting cases: (1) a bare SiO$_2$ slab with no graphene coating ($f=0$), and (2) a SiO$_2$ slab with a full graphene sheet coating ($f=1$). The separation distance is $d = 100$ nm, and the chemical potential is $\mu = 0.5$ eV.

\begin{figure} [htbp]
\centerline {\includegraphics[width=0.5\textwidth]{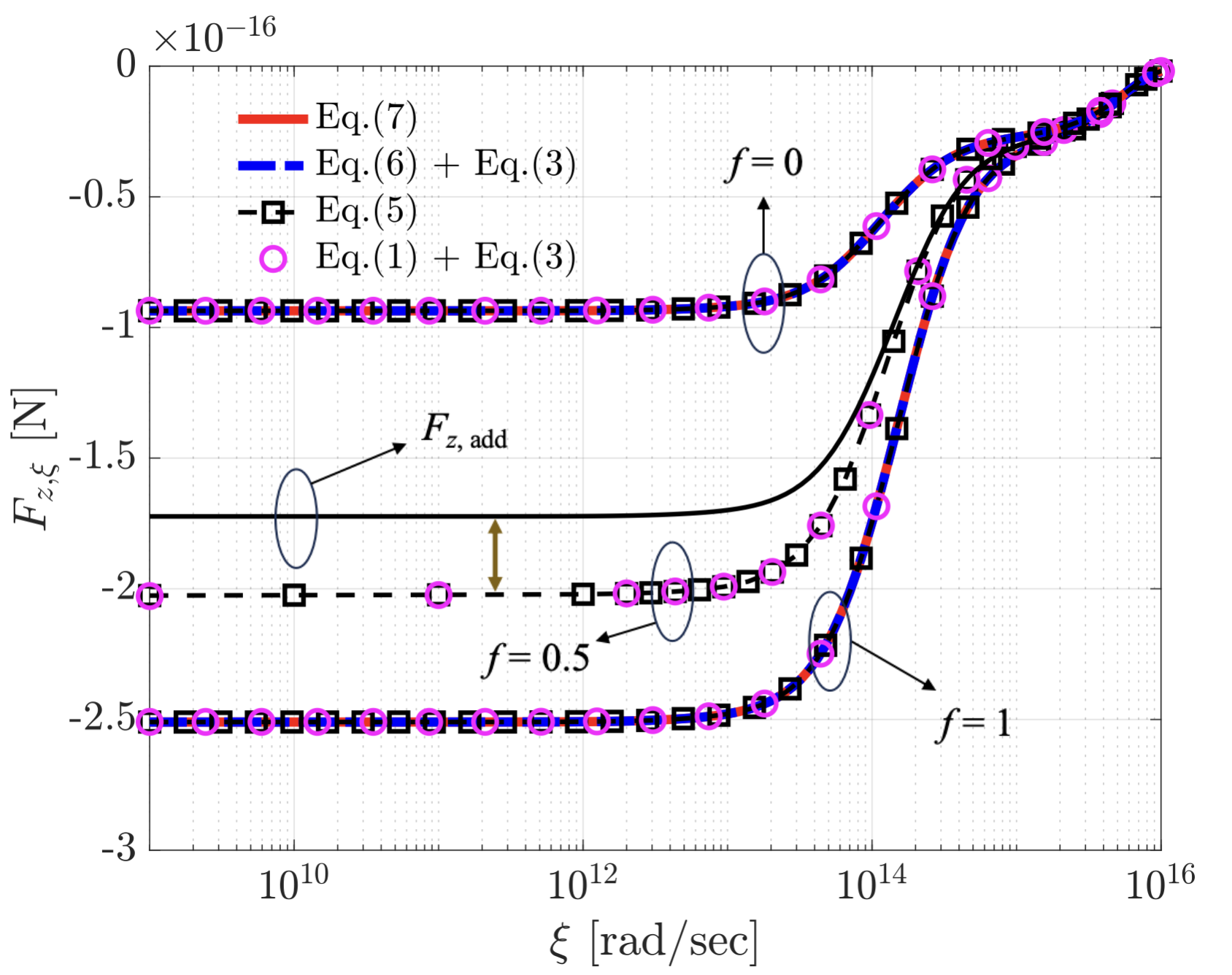}}
\caption{Normal CL force spectra at imaginary frequencies for three configurations: (1) {$f=0$ (bare SiO$_2$ slab without graphene coating)}, (2) {$f=0.5$  (SiO$_2$ slab coated with a half-filled graphene strip grating, and $D=1\;\mu$m)}, and (3) {$f=1$ (SiO$_2$ slab with full graphene coating)}. The black solid line corresponds to the additive approximation $F_{z,{\rm add}}$ (see discussion in the main text). }
\label{fig:Fz_pE_pd_comparison}
\end{figure}

The results are shown in Fig.~\ref{fig:Fz_pE_pd_comparison}, where the spectrum is calculated using two different methods: (i) by performing numerically the partial derivative of the CL energy along the $z$-axis using Eqs.~(\ref{Energy_dipolar}) and ~(\ref{F_z_Der}) , and (ii) by a direct calculation from Eq.~\eqref{CLF}. The equivalence of these two methods is confirmed, and they produce the same normal force spectrum.
In addition to the specific formulas for the grating, for {these} limiting cases, we can use the simpler slab formulas for normal force [Eq.(\ref{CL_Matsubara_slab_particle})] and energy [Eq.(\ref{Energy_slab_particle})]. Again, the numerical equivalence of these {expressions} is demonstrated in Fig.~\ref{fig:Fz_pE_pd_comparison}, showing that for $f=0$ and $f=1$, both methods yield identical results.

At high frequencies ($>10^{15}$rad/s), the normal force spectra of the two configurations ($f=0$ and 1) are close to each other. Nevertheless, at low frequencies ($<10^{13}$rad/s), the spectral normal force of the $f=1$ configuration is considerably greater than that of the $f=0$ configuration, exhibiting a magnitude approximately 2.5 times that of the latter. That is, graphene coating greatly enhances the CL force as compared to the bare SiO$_2$ slab configuration, especially in the low frequency domain.

Next, we consider the case when the slab is covered with a graphene strip grating (filling fraction $f=0.5$ and period $D=1\;\mu$m). We consider here {two} possible approaches to evaluate this force: (i) an exact calculation using the FMM-LBF method, or (ii) a simple and widely used additivity approximation using the relation $F_{z,\text{add}} = f \cdot F_{z}(f=1) + (1-f) \cdot F_{z}(f=0)$. As shown in Fig.~\ref{fig:Fz_pE_pd_comparison}, the normal CL force spectra for these approaches reveal a significant difference. For example, at $\xi = 10^{12}$ rad/s, the difference is about 15\% of the exact value, clearly highlighting a violation of the additive approximation. We also note that, coating the slab with a graphene grating, increases the CL force spectra for small frequencies by about 100\% compared to a bare slab and decreases by about 25\% compared to a slab fully covered with graphene.

\subsection{Effect of the filling fraction $f$ on the Casimir-Lifshitz interactions when the nanoparticle is laterally shifted}  
\begin{figure} [htbp]
\centerline {\includegraphics[width=0.5\textwidth]{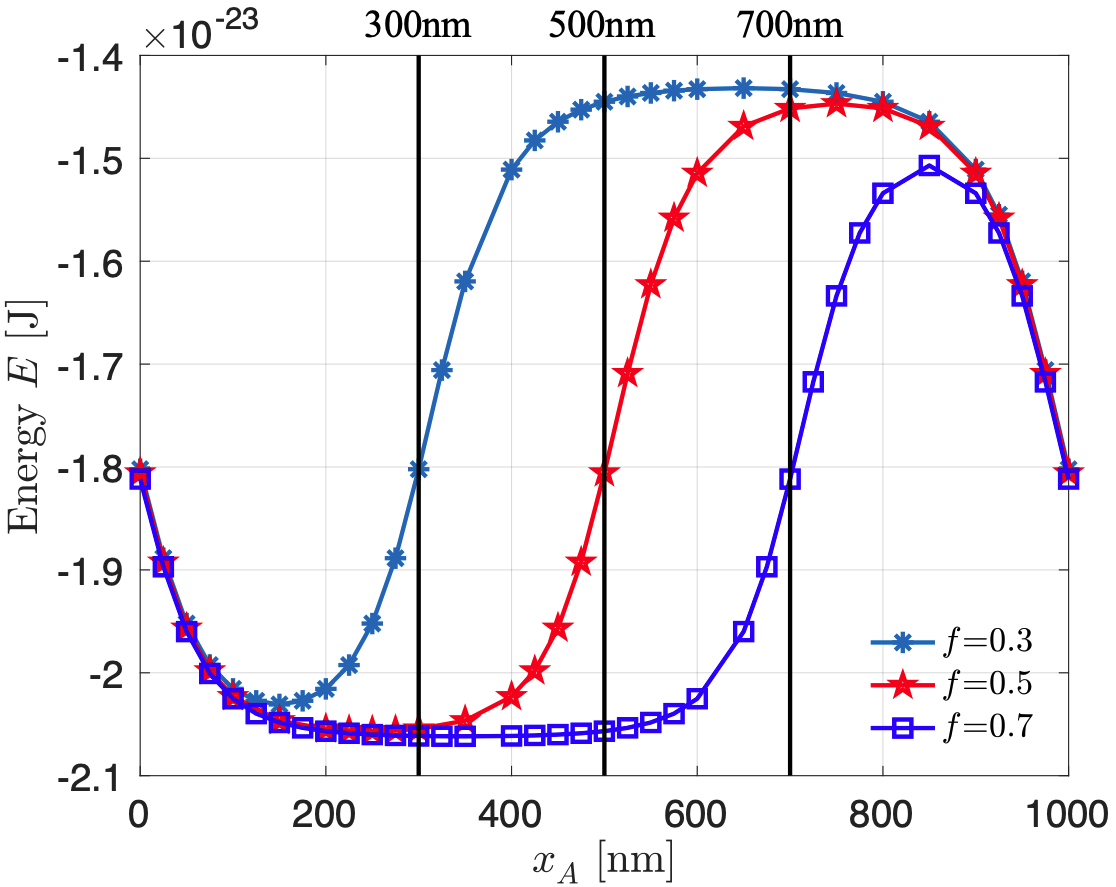}}
\caption{Dependence of CL energy $E$ on the lateral shift $x_A$ for three filling fractions $f = 0.3$, $0.5$, and $0.7$. The vertical lines indicate the positions corresponding to the graphene strip widths for each configuration, i.e., $a = 300$, 500, and 700 nm. Here, $d = 100$ nm, $D = 1\; \mu$m, and $\mu = 0.5$ eV.}
\label{fig:filling_fraction_effect_Energy}
\end{figure}

In this section we explore how the filling fraction $f$ influences the CL interactions when the nanoparticle is laterally shifted by $x_A$. The dependence of the CL energy $E$, calculated using Eq.~(\ref{Energy_dipolar}), on the lateral shift $x_A$ is shown in Fig.~\ref{fig:filling_fraction_effect_Energy} for three different filling fractions: $f = 0.3$, 0.5, and 0.7. Here, we set $d=100$ nm, $D=1\;\mu$m and $\mu=0.5$ eV. The vertical solid black lines at 300 nm, 500 nm and 700 nm correspond to the width $a$ of the graphene strip in the lattice period $D$, for the three different filling fractions, respectively. For the three configurations corresponding to different $f$, we see that, as the nanoparticle laterally shifts above the graphene strip ($0 < x_A < a$) within the spatial period $D$ of the grating, the CL energy decreases to a minimum and then gradually rises. When the nanoparticle is further shifted above the grating slit ($a < x_A < D$), the CL energy increases to a maximum and then falls back. The relative lateral position between the nanoparticle and the graphene grating significantly affects the EM wave scattering between the nanoparticle and the grating, thus influencing the CL energy. Remarkably we see that, in each lattice period, we have a maximum and a minimum in the energy at two different values of the lateral position, corresponding to a stable and unstable lateral equilibrium, respectively. Since the grating is infinitely extending, this corresponds to an infinite series of alternating stable and unstable lateral equilibrium positions. For the specific nanoparticle considered here, the energy well in each period has a barrier height of $\approx  {6  \times 10^{-24}}$ J{, which is the difference between the maximum and the minimum of the energy over one period}. Within the validity range of the point-like dipolar limit description of the nanoparticle we use here, the CL energy is roughly proportional to the static polarizability $\alpha(\omega=0)$ (for SiO$_2$ one has $\varepsilon(0)\approx 4 $ and for $R=20$ nm, the static polarizability is $\alpha(0)\approx 5 {\times} 10^{-23} {\rm m}^{3}$), will scale as $R^3$ for nanoparticles of different radius. The nanoparticle CL energy of Fig.~\ref{fig:filling_fraction_effect_Energy}  can also be easily rescaled to evaluate the CL energy (and force) for an atom, just using the corresponding static atomic polarizability. An atom has a much smaller mass (hence smaller thermal kinetic energy), and can possess a particularly high polarizability, especially in the case of Rydberg atoms.

\begin{figure} [htbp]
\centerline {\includegraphics[width=0.5\textwidth]{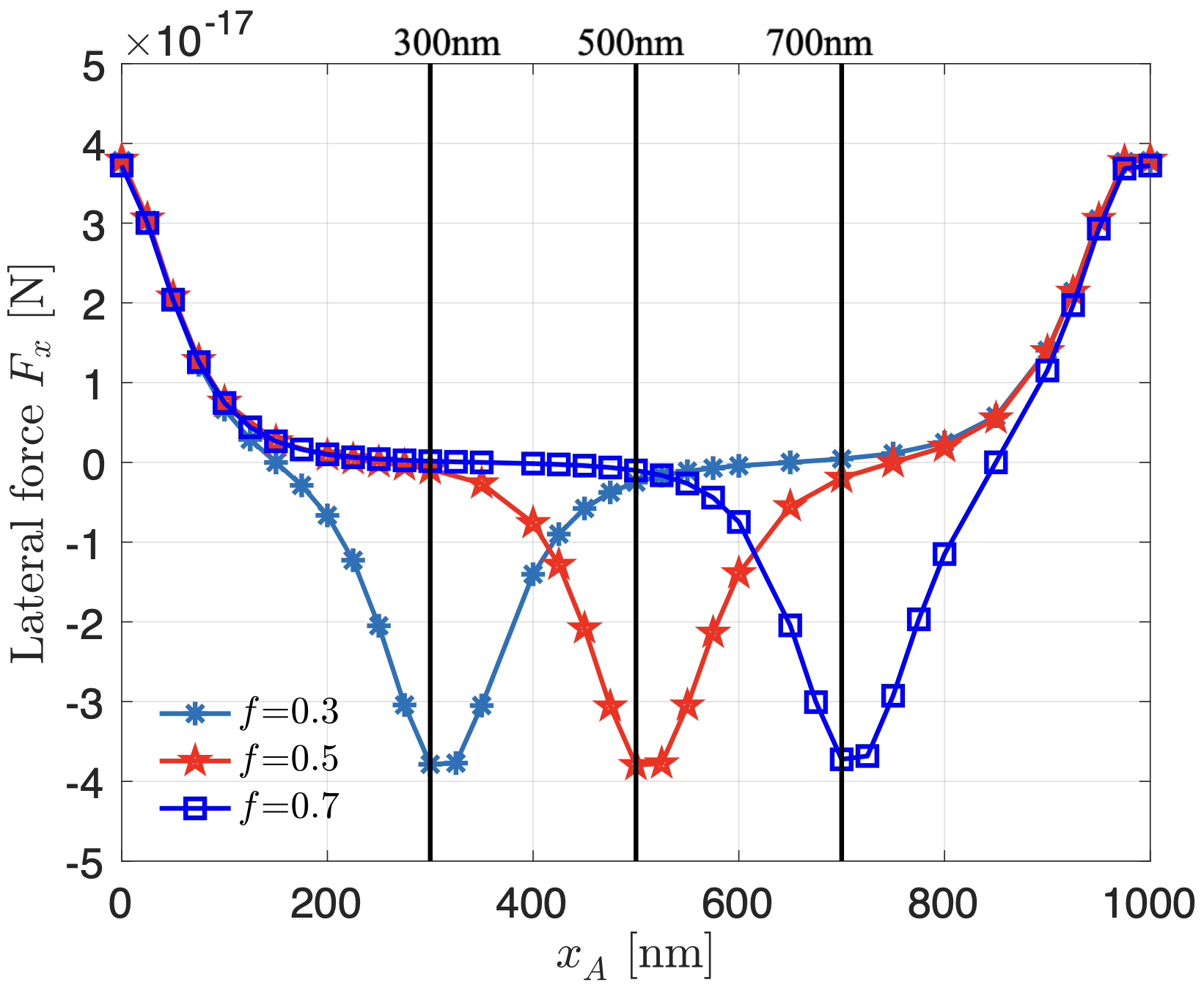}}
\caption{Dependence of the lateral CL force $F_x$ on the lateral shift $x_A$. Same parameters as in Fig.~\ref{fig:filling_fraction_effect_Energy}. }
\label{fig:filling_fraction_effect_Fx}
\end{figure}

Let us now consider the lateral CL force $F_x$ acting on the nanoparticle. It can be calculated from the derivative of the CL energy along the $x$-axis as in Eq.~(\ref{F_x_Der}). In Fig.~\ref{fig:filling_fraction_effect_Fx} we show $F_x$ as a function of the lateral shift $x_A$, and for different values of the filling fraction. Within one grating period, the lateral force $F_x$ acting on the nanoparticle changes sign twice, {first at the graphene strip center ($x_A=0.5a$) and after the slit center [$x_A=0.5(a+D)$].} At these positions, the lateral force cancels, thus marking two mechanical equilibrium positions exactly corresponding to the minimum and maximum of the energy in  Fig.~\ref{fig:filling_fraction_effect_Energy}, respectively. At the first sign change position ($x_A=0.5a$), the nanoparticle force is positive if $x_A < 0.5a$ and  negative $x$ values if $x_A > 0.5a$, implying a lateral stable mechanical equilibrium position. In contrast, the other sign change position [$x_A=0.5(a+D)$] corresponds to an unstable lateral equilibrium.

\begin{figure} [htbp]
\centerline {\includegraphics[width=0.5\textwidth]{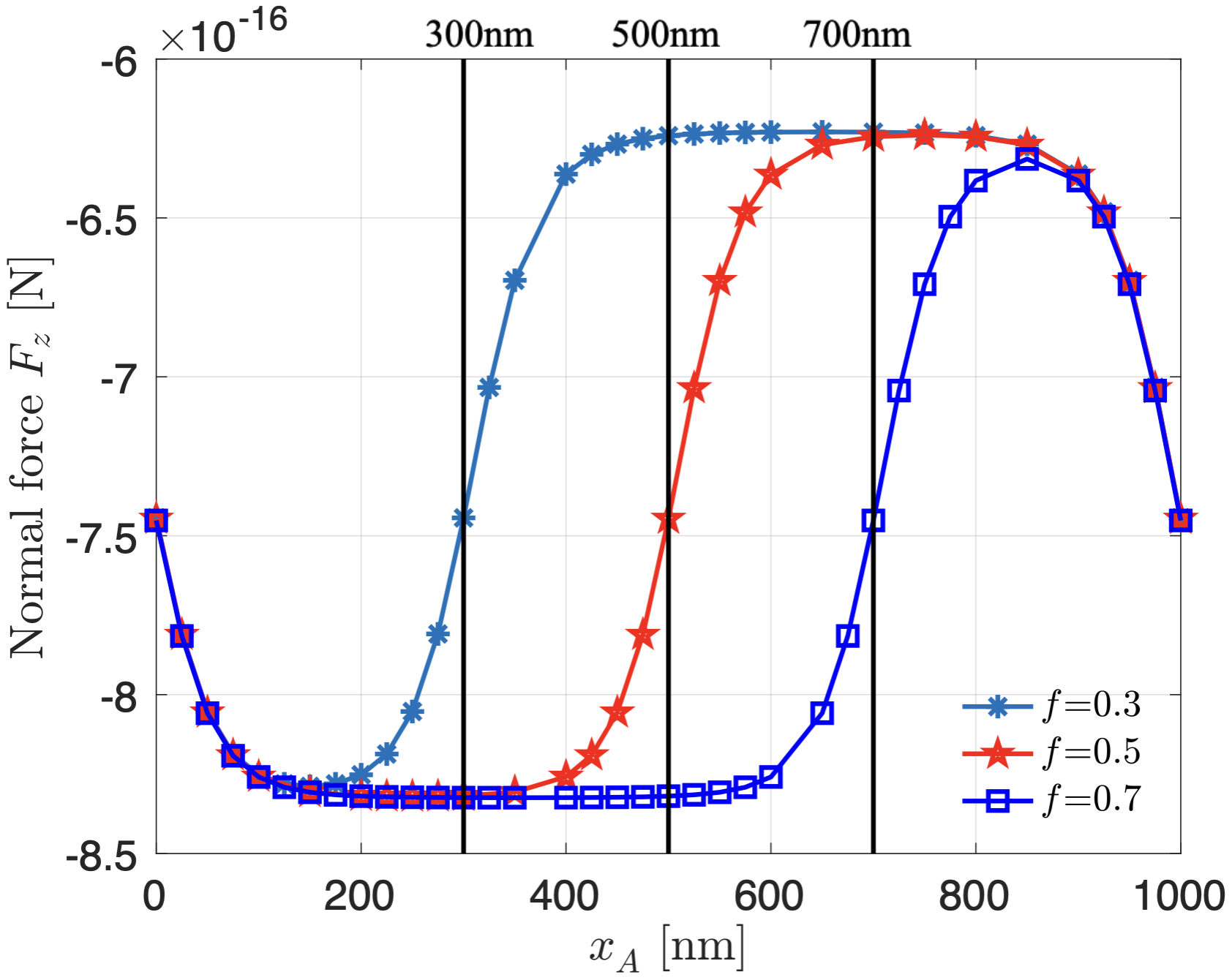}}
\caption{Dependence of the normal CL force $F_z$ on the lateral shift $x_A$. Same parameters as in Fig.~\ref{fig:filling_fraction_effect_Energy}.}
\label{fig:filling_fraction_effect_Fz}
\end{figure}

Let us investigate now the normal CL force and its dependence on the lateral shift $x_A$. The force is calculated through the derivative of the CL energy along the $z$-axis (\ref{F_z_Der}), and the results are shown in Fig.~\ref{fig:filling_fraction_effect_Fz}. The sign of the normal force $F_z$ acting on the nanoparticle remains negative throughout the entire lateral shift range, indicating an attractive force between the two bodies. This force varies with $x_A$ and presents a maximal and a minimal attractive force positions corresponding to the configurations where the nanoparticle is in correspondence of the center of the graphene strip and of the grating slit, respectively. This behavior can be explained since the graphene strips have a metallic behavior, and CL normal force is higher in presence of metallic surfaces since the multiple EM wave reflections are exalted. To quantify the effect of the lateral shift on the force, we define a parameter $\psi = (F_{z\text{max}} - F_{z\text{min}})/F_{z\text{min}}$. While the dependence of $F_z$ on $x_A$ varies for the three configurations corresponding to different filling fractions, it exhibits a similar $\psi$ value of approximately 31\%.

\subsection{Effect of the lateral shift $x_A$ on the Casimir-Lifshitz interactions at different separation distances $d$}
\begin{figure*} [htbp]
\centerline {\includegraphics[width=1.\textwidth]{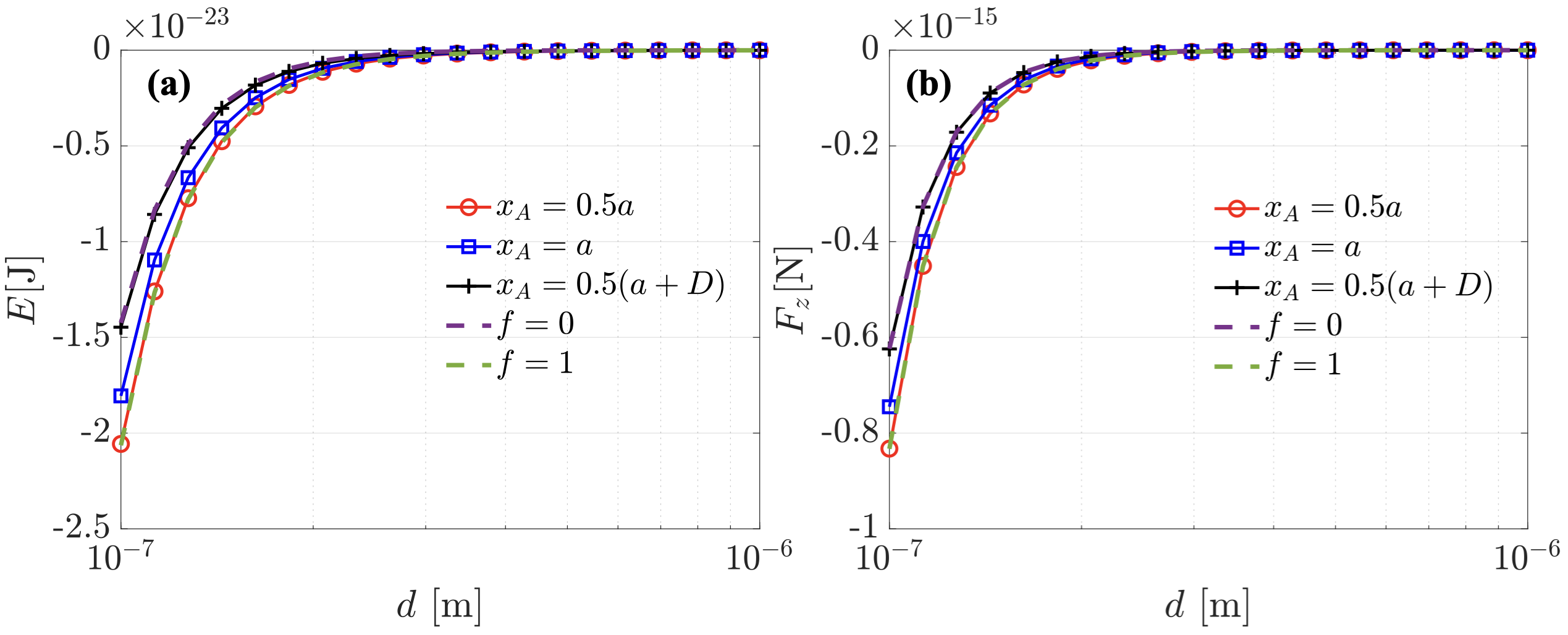}}
\caption{Dependence of (a) CL energy and (b) Normal CL force on the separation distance $d$. Here, three lateral shifts are considered, $x_A=0.5a$, $a$ and $0.5(a+D)$, $\mu=0.5$eV, and $D=1\mu$m.}
\label{fig:separation_dependence}
\end{figure*}

In the previous section, the CL energy and forces have been evaluated for a fixed value of the separation distance $d$, as a function of the lateral shift. Here we consider the complementary investigation. In Fig.~\ref{fig:separation_dependence} we show the dependence of both the CL energy and normal CL force on the separation distance $d$, for three different values of lateral shift: $x_A=0.5a$ (i.e. the center of the graphene strip), $a$ (i.e. the edge of the graphene strip), and $0.5(a+D)$ (i.e the center of the grating slit). Here $f=0.5$, $\mu=0.5$ eV, and $D=1\;\mu$m. As the separation distance increases, both the CL energy and the normal force exhibit a monotonic variation. At short separations, we can observe notable differences between the three configurations. This is less and less the case, as the separation distance increases.

At short separations ($\lesssim 300$  nm), when the nanoparticle laterally shifts from the graphene strip center ($x_A=0.5a$) to the strip edge ($x_A=a$), and further to the slit center [$x_A=0.5(a+D)$], it is more sensitive to the local metallic-dielectric change in the material located beneath it. This local interaction accounts for different scattering behaviors in the three configurations and thus affects significantly the force. As shown in Fig.~\ref{fig:separation_dependence} (b), the normal CL force of the configuration with $x_A=0.5a$ is the highest among the three configurations considered here, where the particle feels a graphene coating, resulting in an enhancement of the force. When the nanoparticle shifts to the center of the slit in the configuration of $x_A=0.5(a+D)$, the particle no longer feels the graphene coating and, consequently, the force is the lowest of the three configurations. When the separation distance increases, the nanoparticle {feels} the graphene strips and the grating slits as an effective and homogeneous whole, rather than a grating with geometric details, hence the effect of the lateral shift is less influent.

In the same Fig.~\ref{fig:separation_dependence}, we report the cases of $f=1$ and $f=0$, and clearly observe that, at short separation distances, the energy and the normal force, when the nanoparticle is at the center of the graphene strip (respectively at the center of the slit), are very similar to those for $f=1$ (respectively $f=0$). This is coherent with the local interaction effect explained above. Besides, when the nanoparticle is positioned at the edge of the strip ($x_A=a$), the energy and force values fall between those for $f=0$ and $f=1$. 

\subsection{Effect of the chemical potential $\mu$ on the Casimir-Lifshitz interactions with lateral shift}
\begin{figure} [htbp]
\centerline {\includegraphics[width=0.5\textwidth]{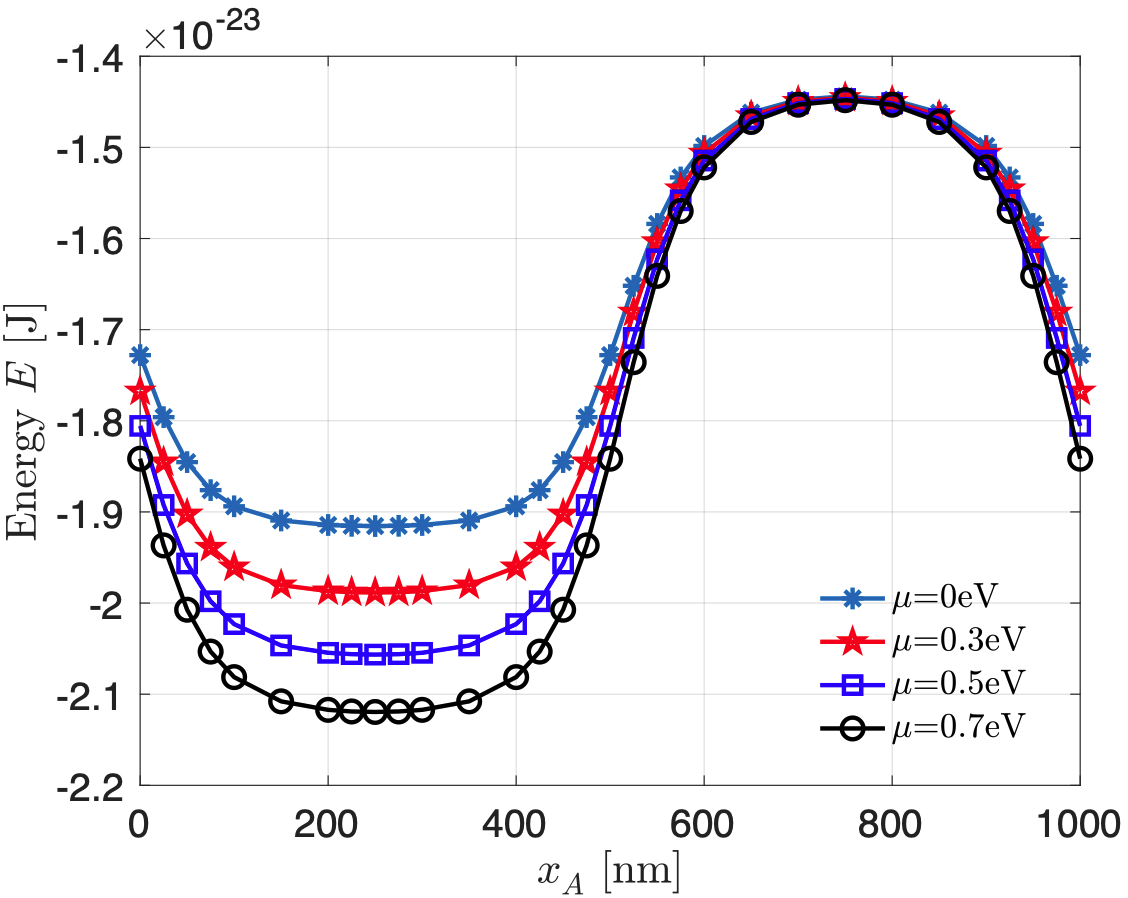}}
\caption{Dependence of the CL energy $E$ on the lateral shift $x_A$. Four different chemical potential are considered, $\mu=0.$, $0.3$, $0.5$ and $0.7$eV. Separation distance $d=100$nm, grating period $D=1\mu$m, filling fraction $f=0.5$.}
\label{fig:chemical_potential_effect_Energy}
\end{figure}

We move now to {investigate} the modulation of CL interactions by variyng the chemical potential $\mu$ of the graphene.
In this section, four different chemical potentials are considered, $\mu=0$, 0.3, 0.5 and 0.7 eV. As for the geometric parameters, we set $d=100$ nm, $D=1\; \mu$m and $f=0.5$. The dependence of the CL energy $E$ on the lateral shift $x_A$ is shown in Fig.~\ref{fig:chemical_potential_effect_Energy}.

\begin{figure} [htbp]
\centerline {\includegraphics[width=0.5\textwidth]{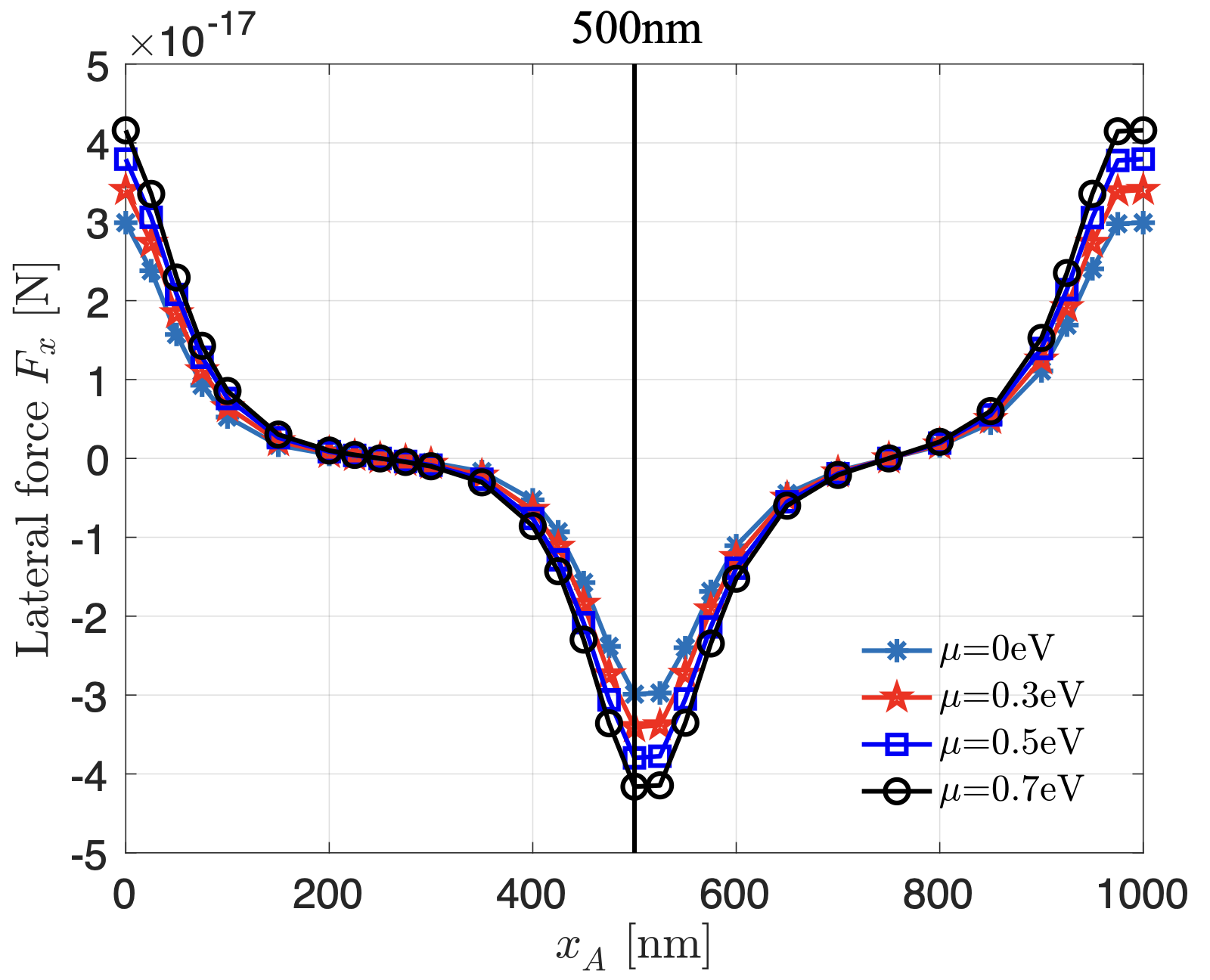}}
\caption{Dependence of the lateral CL force $F_x$ on the lateral shift $x_A$. The vertical line is for the lateral shift $x_A$ equal to the width of the graphene strip. Same parameters as in Fig.~\ref{fig:chemical_potential_effect_Energy}.}
\label{fig:chemical_potential_effect_Fx}
\end{figure}

In this figure, we can identify two distinct regions: (1) the graphene strip region ($0 < x_A < a$), where the chemical potential $\mu$ significantly affects the dependence of the CL energy on the lateral shift $x_A$; and (2) the slit region ($a < x_A < D$), where the dependence of the CL energy on $x_A$ remains similar across the considered configurations. For the former, when the nanoparticle is above the graphene strip, the scattering details are strongly linked to the optical properties of the graphene coating, which is typically characterized by $\mu$. Thus, in this region, manipulating the chemical potential of the graphene will considerably affect the scattering details, thereby modulating the CL interaction. For the latter, the effect of the chemical potential becomes less significant, especially when the nanoparticle is above the center of the slit. This is because, in this region, there is no graphene coating. Therefore, the significant impact of the chemical potential on the scattering details is not equivalent for any position in the grating period, and is influenced by the proximity of nanoparticle and the graphene strips. 

Regarding the lateral CL force, we show its dependence on the lateral shift $x_A$ in Fig.~\ref{fig:chemical_potential_effect_Fx}. When the nanoparticle approaches the center of the graphene strip or the center of the slit, regardless of the graphene's chemical potential $\mu$, this force always approaches zero. However, when the nanoparticle reaches one of the edges of graphene, the chemical potential effect becomes significant. Particularly, we can see clear differences among the four configurations of different $\mu$ when the nanoparticle is located at $x_A=$ 0, $a$ and $D$. The lateral force itself is induced by the asymmetry of the structure in the lateral direction. The maximum of the absolute value of the lateral shift force is obtained when the nanoparticle approaches the graphene strip edges. 

\subsection{Different regimes for the normal and lateral shift effects as a function of $d/D$}
As illustrated in Fig.~\ref{fig:filling_fraction_effect_Energy}, Fig.~\ref{fig:filling_fraction_effect_Fz}  and Fig.~\ref{fig:chemical_potential_effect_Energy}, we see that moving the nanoparticle toward the graphene strip centers ($x_A=a/2$) or slit centers ($x_A=(D+a)/2$) results in a notable lateral shift effect on CL interactions. 

\begin{figure} [htbp]
\centerline {\includegraphics[width=0.5\textwidth]{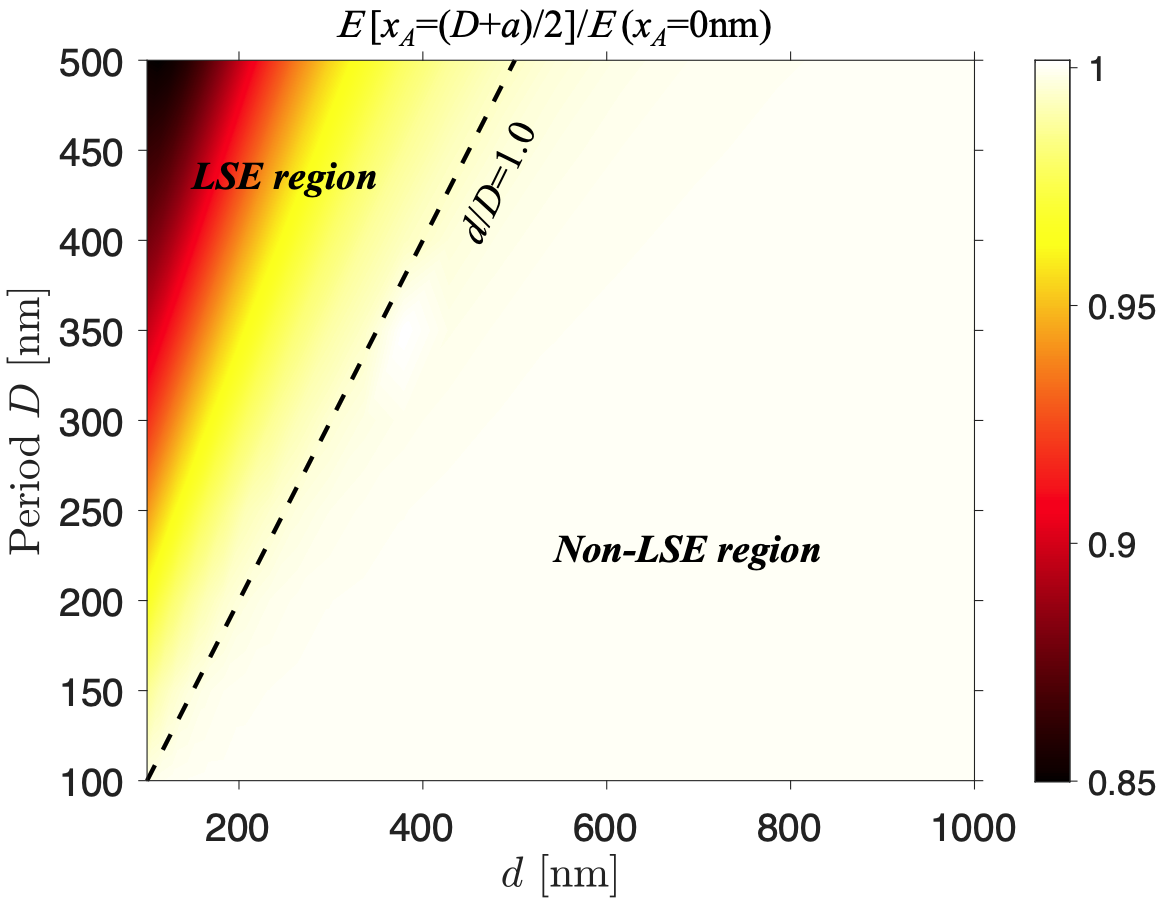}}
\caption{Dependence of the ratio $\varphi = E[x_A=(D+a)/2]/E(x_A=0{\rm nm})$ on the normal shift $d$ and period $D$ for $f=0.5$ and $\mu=0.5$ eV. The line (geometric factor $d/D = 1$) is added for reference.}
\label{fig:regime_Energy}
\end{figure}

It is interesting now to explore the values of the lattice period $D$ and separation $d$ for which a maximal variation of energy as a function of the lateral shift is available, in each lattice period. We hence define the ratio {$\varphi =E[x_A=(D+a)/2] / E(x_A=0 {\rm nm})$} to evaluate the lateral shift effect on the CL interaction. In order to get such lateral shift regime, we show the dependence of the ratio $\varphi $ on the separation $d$ and period $D$ in Fig. \ref{fig:regime_Energy} for $f=0.5$ and $\mu=0.5$ eV. The line (geometric factor $d/D = 1.0$) is added for reference.

As shown in Fig.~\ref{fig:regime_Energy}, there are two distinct regions: (1) a Lateral Shift Effect (LSE) region above the auxiliary line $d/D=1$, and (2) a non-LSE region below this line. In the LSE region, where the geometric factor $d/D<1$, the ratio $\varphi$ significantly deviates from 1 (much less than 1). This indicates that the lateral shift significantly reduces the CL energy when the nanoparticle is positioned above the grating slit. When the separation is less than the grating period, the scattering details are significantly influenced by the relative positions of the two bodies. Therefore, a lateral shift will have a notable impact on the scattering, and consequently, on the CL interactions.
In the non-LSE region, where the geometric factor $d/D>1$, the ratio $\varphi$ is approximately 1, and the lateral shift effect is negligible. Due to the relatively large separation compared to the grating period, the nanoparticle sees the graphene grating as an effective and homogeneous whole. As a result, the lateral shift of the nanoparticle does not influence the scattering details, provided that the separation remains constant. Consequently, the lateral shift effect on the CL interactions becomes less significant.

\section{conclusion}

We studied the effect of normal and lateral shifts on Casimir-Lifshitz interactions (energy, normal force and lateral force) between a nanoparticle and a finite-thickness planar fused silica slab coated with a graphene grating. We show that the frequency spectrum of the interaction is strongly affected by the presence of the grating covering, in particular at low frequencies. The CL energy presents a considerable variation with respect to the lateral nanoparticle displacement. In particular the CL energy presents a series of alternating minima (lateral stable equilibrium) and maxima (lateral unstable equilibrium), in correspondence of the center of the graphene strips and of the grating slits, respectively. By varying the graphene filling fraction, the relative positions of these two equilibrium position can be changed. The lateral force behaves accordingly, and show regions of alternating sign. We see that the normal CL force also has a strong variation and shows an alternating maximum and a minimum {in each lattice period}, still remaining always attractive.

The effect of the variation of the graphene chemical potential on the CL  energy is not uniform. It is higher in the center of the graphene strips, while it is almost negligible in the grating slit centers.  For a fixed graphene grating period ($D=1\mu$m) and not too large normal shift (separation $d<400$nm), the lateral shift effect on the CL force is significant. When increasing further the separation $d$, the lateral shift effect becomes less important. 

Finally, we show that the lateral shift effect is sensitive to the geometric factor $d/D$, and the line $d=D$ separates two different regions, where the effect of the lateral shift is relevant ($d<D$) or negligible ($d>D$), respectively.

Our findings can be relevant for the study of the interaction of nanoparticles with nano/micro mechanical devices, and their possible manipulation at the nanoscale. The manipulation can {be} done by varying the graphene filling fraction, and/or \emph{in situ} by varying the graphene chemical potential through the application of a simple gate voltage. Since these effects are easily rescaled with the nanoparticle polarizability, our results are directly valid, with a simple scaling factor, in the case of atoms interacting with the same nanostructures.

\begin{acknowledgments}
This work described was supported by a grant "CAT" (No. A-HKUST604/20) from the ANR/RGC Joint Research Scheme sponsored by the French National Research Agency (ANR) and the Research Grants Council (RGC) of the Hong Kong Special Administrative Region, China.
\end{acknowledgments}

\bibliography{Force_Energy}

\end{document}